\documentclass[cameraready]{Interspeech}
\usepackage{amssymb,multirow,cite}

\title{Vo{X}tream2: Full-stream TTS with dynamic speaking rate control}

\author[affiliation={1}, orcid=0009-0009-5292-1737]{Nikita}{Torgashov}
\author[affiliation={1}, orcid=0000-0002-1643-1054]{Gustav Eje}{Henter}
\author[affiliation={1}, orcid=0000-0002-8579-1790]{Gabriel}{Skantze}

\address{
    $^1$ Department of Speech, Music and Hearing, KTH Royal Institute of Technology, Sweden
}

\email{nikitat@kth.se, ghe@kth.se, skantze@kth.se}


\usepackage{comment}

\begin{document}

\maketitle

\begin{abstract}

Full-stream text-to-speech (TTS) for interactive systems must start speaking with minimal delay while remaining controllable as text arrives incrementally. We present VoXtream2, a zero-shot full-stream TTS model with dynamic speaking-rate control that can be updated mid-utterance on the fly. VoXtream2 combines a distribution matching mechanism over duration states with classifier-free guidance across conditioning signals to improve controllability and synthesis quality. Prompt-text masking enables textless audio prompting, removing the need for prompt transcription. Across standard zero-shot benchmarks and a dedicated speaking-rate test set, VoXtream2 achieves competitive objective and subjective results against public baselines despite a smaller model and less training data. In full-stream mode, it runs 4 times faster than real time with 74\,ms first-packet latency on a consumer GPU.

\end{abstract}

\section{Introduction}

Recent progress in neural text-to-speech (TTS) synthesis has led to highly natural and intelligible speech generation. However, most contemporary systems implicitly assume that speaking rate is static across an utterance, typically allowing only coarse, global control over speed. This assumption contrasts sharply with human speech behavior. In spontaneous communication, speaking rate is inherently dynamic: speakers slow down when formulating thoughts, insert pauses and filler words such as ``uhm" or ``hmm", and accelerate when expressing well-prepared or information-dense content. These fluctuations often occur within a single sentence, reflecting cognitive load, discourse structure, and communicative intent. The absence of such fine-grained temporal variation in current TTS systems results in speech that, while clear and fluent, lacks the spontaneity and realism characteristic of natural human interaction.

\footnotetext{Under review at INTERSPEECH'26.}

It has long been argued that conversational agents must be able to generate speech incrementally \cite{schlangen_general_2009,skantze_towards_2013}. 
In recent years, the growing adoption of voice-driven interfaces based on large language models (LLMs) has amplified the need for streaming TTS systems capable of operating under strict latency constraints \cite{shikhar2025llmvox}. In emerging applications such as real-time conversational agents and speech-to-speech translation, text is generated incrementally and must be converted into speech on the fly \cite{bai2025speakstream}. To support seamless interaction, TTS systems must process streaming text input and produce short waveform chunks with minimal delay. Despite recent advances, most existing TTS architectures remain fundamentally offline \cite{chen2025neural,ju2024naturalspeech,mehta2024matcha}, requiring full-utterance text before synthesis. Among the limited number of streaming-capable systems, controllability, particularly over speaking rate, is either absent or constant on the utterance level.

\begin{table}[t]
  \caption{Comparison of VoXtream2 with other speaking rate control models. Spark-TTS-Ctrl stands for controllable version.}
  \label{tab:features_comparison}
  \centering
  \hspace*{-0.1cm}%
  \resizebox{0.47\textwidth}{!}{
      \begin{tabular}{lccccc}
        \toprule
        \multirow{2}{*}{\textbf{Model}} &
        \textbf{Open} &
        \textbf{Voice} &
        \textbf{Full} &
        \textbf{Dynamic} \\
        &
        \textbf{Weight} &
        \textbf{Cloning} &
        \textbf{Stream} &
        \textbf{Control}
        \\
        \midrule
        Parler-TTS \cite{lyth2024natural}        & \checkmark &            &            &            \\
        Spark-TTS-Ctrl \cite{wang2025spark}       & \checkmark &            &            &            \\
        ControlSpeech \cite{ji2025controlspeech} &            & \checkmark &            &            \\
        MaskGCT \cite{wang2024maskgct}           & \checkmark & \checkmark &            &            \\
        VoiceStar \cite{peng2025voicestar}       & \checkmark & \checkmark &            &            \\
        WeSCon \cite{wang2025word}               &            & \checkmark &            & \checkmark \\
        CosyVoice2 \cite{du2024cosyvoice2}       & \checkmark & \checkmark & \checkmark &            \\
        VoXtream2 (Ours)                         & \checkmark & \checkmark & \checkmark & \checkmark \\
        \bottomrule
      \end{tabular}
    }
    \vspace{-4ex}
\end{table}

These two limitations, static speaking rate modeling and non-streaming generation, are tightly coupled in practice. Real-time conversational speech demands not only low latency but also adaptive prosody that reflects the evolving state of the dialogue. As text unfolds token by token, the appropriate speaking rate may change dynamically, mirroring the uncertainty, emphasis, or urgency conveyed by the underlying language model. 

To bridge the gap between highly intelligible synthetic speech and truly human-like spoken interaction, we introduce VoXtream2 - a full-stream zero-shot TTS with dynamic speaking rate control (SRC). The model utilizes distribution matching and Classifier-free guidance (CFG) for a fine-grained SRC on the frame level, which can be adjusted as the model generates speech. For static SRC, our model shows high stability in terms of intelligibility and voice cloning across various speaking rates, and also enables a dynamic SRC in the stream. The model works four times faster than real-time and achieves 74 milliseconds of initial latency in full-stream mode on the user-grade GPU, allowing for seamless interaction. Table \ref{tab:features_comparison} highlights these capabilities and compares our model with prior works. Our main contributions are as follows:

\begin{itemize}
  \item We introduce a dynamic speaking-rate control mechanism in the stream, which can be modified on the fly.
  \item We leverage CFG not only for quality improvement, as was done in previous works, but also investigate its impact on speech rate control.
  \item By utilizing prompt text masking, we decrease the reliance on acoustic prompt speed and content, making the system more practical and accurate.
  \item The performance of our model on zero-shot TTS is on par with state-of-the-art systems, even though it uses less training data and is among the smallest compared to competitors.
\end{itemize}
Audio examples, pretrained model, and code are available at \url{https://herimor.github.io/voxtream2/}.

\section{Related Work}

\subsection{Speaking Rate Control}
\textbf{Static control}. Speaking rate control in TTS has been extensively explored through duration modeling. Early approaches introduce SRC by conditioning end-to-end models on sentence-level rate descriptors \cite{bae20_interspeech} or by manipulating encoder embeddings to bias alignments, effectively modifying speech pace via implicit duration changes \cite{lenglet22_interspeech}. Subsequent works formalize SRC in non-autoregressive frameworks by conditioning or modifying duration predictors \cite{bandekar2023speaking}, enabling sentence-level duration scaling via global duration control \cite{lee24m_interspeech}, word or phoneme-level duration scaling through duration predictor manipulation \cite{kim2024masked}, or using attention-based or probabilistic duration modeling \cite{ogura2025phoneme}. Although these methods achieve fine-grained temporal control, they are typically evaluated on single-speaker or limited multi-speaker datasets and do not address modern TTS requirements such as zero-shot voice cloning or streaming synthesis.

A large number of recent TTS systems investigate the control of speaking rate by proposing fundamentally different approaches that can be divided into three main categories. The first category \cite{guo2023prompttts,shimizu2024prompttts++,leng2024prompttts2,lyth2024natural,hu2026voicesculptorvoicedesigned} uses a detailed text description of the target voice, including the speaking rate. These works show a high accuracy of the SRC. However, they do not include a voice cloning capability. The second category \cite{yang2024instructtts,liu23t_interspeech,ji2025controlspeech,du2024cosyvoice2} utilizes acoustic prompt and text instructions as model input, thereby enabling voice cloning with control of speaking style. Another work \cite{wang2025spark} overlaps with both categories by enabling voice cloning and category-based gender, pitch, and SRC in a single model. However, the model can only do either voice cloning or controlled generation. The third group \cite{wang2024maskgct,chen2025f5,peng2025voicestar,zhou2025indextts2} controls the target duration of the generated speech, which can be utilized to control speaking rate. Even though models from the last two categories can do both voice cloning and manipulate the speaking rate, they only enable utterance-level SRC.

\textbf{Dynamic control}. One of the latest works \cite{wang2025word} introduces dynamic SRC at the word-level by resampling input speech tokens. Even though the proposed method does not require extensive model retraining, it relies on a complex multi-round inference. The SRC does not scale well, as downsampling by more than 40\% of the original input length leads to unintelligible results. Overall, dynamic SRC seems to be underexplored, as we only found a single relevant paper applying it to modern TTS. 

\subsection{Full-Stream TTS}

A full-stream TTS system is an architecture that operates entirely in streaming mode, accepting incrementally generated text tokens as input and producing small waveform chunks as output. A full-stream system begins generating speech before the full text is available, minimizing first packet latency. 

Modern zero-shot full-stream TTS models can work faster than real-time and produce high-quality speech with limited look-ahead. Some models \cite{yang2024interleaved,du2024cosyvoice2} interleave text and speech tokens, others \cite{sheng2025syncspeech} use a temporal-masked transformer, or utilize monotonic alignment \cite{torgashov2026voxtream} between phonemes and audio frames. However, all these models require text transcription of the acoustic prompt, making their usage less practical. For example, acoustic prompts with a high speaking rate are difficult to transcribe and align, which might result in an incorrect initial state. One of the most recent works \cite{kyutai2025streaming} does not require text transcription of the prompt, but delays the audio output stream to accumulate input text context, resulting in high initial latency.

\subsection{Classifier Free Guidance} 

Classifier-free guidance was initially introduced for diffusion-based image generative models \cite{ho2021classifierfree} and was widely adopted for non-autoregressive (NAR) \cite{le2023voicebox,wang2024maskgct,chen2025f5,zhu2025zipvoice} and autoregressive (AR) \cite{kreuk2023audiogen,darefsky2024parakeet,wang2025ssr,hussain2025koel,kyutai2025streaming} audio generative models. The authors of \cite{hussain2025koel} showed that in zero-shot TTS, CFG can be applied not only to the text condition, but also to the audio, significantly improving the speaker similarity of the produced speech. Parakeet \cite{darefsky2024parakeet}, and SSR-Speech \cite{wang2025ssr} noticed that applying CFG leads to an accelerated pace of the generated speech. In this work, we show how SRC can benefit from the speed-up property of CFG.

\subsection{Distribution matching}

Distribution matching has been explored in image generation, through differentiable histogram losses for image-to-image translation \cite{aharon2023huenet} and exact feature distribution alignment for style transfer \cite{zhang2022efdm}. In text generation, related ideas appear as probability reweighting during autoregressive decoding \cite{yang2021fudge} and multiplicative likelihood adjustment for attribute control \cite{krause2020gedi}. In contrast, histogram-level distribution matching with online correction has received little attention in controllable speech generation, particularly for dynamic SRC.

\section{Method}

\begin{figure}[t]
  \centering
  \includegraphics[width=\linewidth]{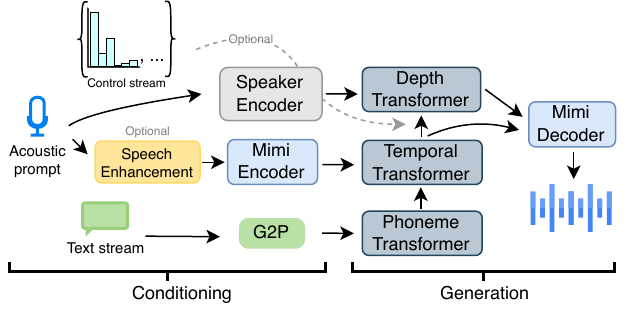}
  \caption{Overview of VoXtream2 architecture.}
  \label{fig:arch}
  \vspace{-1ex}
\end{figure}


We build on our earlier VoXtream model \cite{torgashov2026voxtream}, which is among the fastest full-stream TTS architectures, making it a strong foundation for the developments introduced in the present work. However, this model has several limitations, which we address in this work, and also lacks an SRC mechanism.

\subsection{Model architecture}

Our model architecture follows the one proposed in VoXtream with some modifications. The overview of VoXtream2 architecture is presented in Figure \ref{fig:arch}. For the encoding of phoneme sequences, we used the incremental Phoneme Transformer (PT). Unlike the baseline model, we used the International Phonetic Alphabet (IPA) dictionary, opening up for multilingual generation \cite{casanova24_interspeech}, and increased the maximum look-ahead size to 25 phonemes to achieve better prosody. The model requires at least 3 phonemes of minimum look-ahead to keep the full-stream generation intelligible. The main generative component is an autoregressive Temporal Transformer (TT), which predicts semantic and duration tokens. It is conditioned on the outputs of PT and audio tokens extracted by the Mimi \cite{defossez2024moshi} codec, operating at 12.5 Hz. The phoneme embeddings are assigned to the corresponding audio frames via monotonic alignment. Compared to the original VoXtream model, we increased the number of duration tokens to achieve a more fine-grained duration control. At each step, the model predicts the shift state, indicating how many phonemes to advance, from the range $[0,2]$, and a number of phonemes per frame, which is either 1 or 2, resulting in 6 duration tokens. The autoregressive Depth Transformer (DT), conditioned on the output embedding of TT, speaker embedding, and a semantic token, is used to predict acoustic tokens of Mimi. Similar to \cite{kyutai2025streaming}, we used 16 codebooks for a better speech quality. To enable punctuation handling, missing in VoXtream, we added each punctuation symbol as a separate phoneme token to the PT and then removed the corresponding outputs before feeding them to TT. This way, TT only learns to model duration for phoneme tokens but also gets the contextual information about punctuation in the input sequence.

The number of outputs of TT is defined as the Mimi vocabulary size $N$ multiplied by 6 duration tokens, so that semantic tokens and duration tokens are modeled jointly, and optimized via cross-entropy loss. The DT outputs 15 acoustic tokens at each step and is optimized via the same loss function. The final optimization objective is defined as a sum of TT and DT losses.

\subsection{Prompt Text Masking}
\label{sec:prompt_masking}

The original VoXtream model requires text transcription of the acoustic prompt and relies on an external phoneme aligner. This provides paired audio-text examples that help the model more accurately mimic the target voice. However, errors introduced by the aligner may degrade performance. To address this issue, we remove the text from the prompt, forcing the model to rely solely on the audio signal. Our approach is similar to \cite{liu2026cross}. However, instead of dropping the text prefix, we replace it with special tokens and include them in gradient computation, which enables CFG. During training, we randomly select the first 3 to 10 seconds of the audio and mask the corresponding text tokens with a sequence of \texttt{<UNK>} tokens. During generation, we assign a single \texttt{<UNK>} token to each acoustic frame of the prompt.

Similar to \cite{liu2026cross}, prompt masking enables translingual capabilities (any language to English). Since this is not the primary focus of our work, we do not report quantitative evaluations but provide corresponding audio examples on the demo page.

\subsection{Classifier Free Guidance}

Following \cite{hussain2025koel}, we applied CFG to every conditioning in our model. During training, we mask around 10\% of the text tokens in the prefix of every sequence (Section \ref{sec:prompt_masking}), on the input of PT. We also mask the corresponding audio tokens in the prefix with a probability of 10\% on the input of TT. The speaker embedding used as an extra conditioning of the DT is dropped with a probability of 10\%. We used different $\gamma$ values for TT and DT logits. After analysis of experiments with the guidance scale at Koel-TTS \cite{hussain2025koel}, we used $\gamma_{temp}=1.5$ in the TT to allow for more prosodic variation. For the DT, we used $\gamma_{depth}=3.0$ for a more precise control over the target speaker's voice. To further improve the voice cloning, we increased the weight of the speaker embedding conditioning by 50\%.

\subsection{Acoustic Prompt Enhancement}

After applying CFG, we observe that increasing speaker similarity degrades signal quality, which we monitor using the UTMOS metric \cite{saeki22c_interspeech}. As the model more closely mimics the target voice, the synthesized speech quality converges to that of the acoustic prompt. Therefore, if the prompt contains background noise or recording artifacts, these imperfections may propagate to the generated audio, especially at high $\gamma_{depth}$ values. To mitigate this effect, we enhance the acoustic prompt using the Sidon model \cite{nakata2026sidon}, following \cite{giraldo2026zero}. Since enhancement is applied only to the prompt, it does not increase generation latency and helps preserve signal quality under CFG.

\begin{table*}[t]
\caption{Evaluation results of zero-shot TTS models. Bold for the \textbf{best} result, underline for the \underline{second-best} result. Naturalness is measured as MUSHRA scores (0--100). VoXtream2-B denotes baseline model without acoustic prompt enhancement.}
\label{tab:eval}
\centering
\footnotesize
\setlength{\tabcolsep}{6pt}

\resizebox{1.0\textwidth}{!}{
\renewcommand{\arraystretch}{1.15}
\begin{tabular}{@{}l*{10}{c}@{}}
\toprule
  \multirow{2}{*}{\textbf{Model}} &
  \multirow{2}{*}{\textbf{\#Data(h)}} &
  \multirow{2}{*}{\textbf{\#Params}} &
  \multicolumn{3}{c}{\textbf{SEED \textit{test-en}}} &
  &
  \multicolumn{3}{c}{\textbf{LibriSpeech-PC \textit{test-clean}}} &
  \textbf{Naturalness $\uparrow$} \\ 
  \cline{4-6}
  \cline{8-10}
  & 
  & &
  \textbf{WER (\%) $\downarrow$} &
  \textbf{SPK-SIM $\uparrow$} &
  \textbf{UTMOS $\uparrow$} &
  &
  \textbf{WER (\%) $\downarrow$} &
  \textbf{SPK-SIM $\uparrow$} &
  \textbf{UTMOS $\uparrow$} &
  \textbf{$\boldsymbol{\mu}$ $\pm$ 95\% CI} \\
\midrule
Human      & -           & -     & 1.50             & 0.734             & 3.53             & & 1.95             & 0.695             & 4.10             & 56.7 $\pm$ 2.6             \\
\midrule
CosyVoice2 & 167k Multi. & 618M  & 2.27             & 0.658             & \textbf{4.16}    & & \underline{1.93} & 0.655             & \textbf{4.38}    & \underline{66.2} $\pm$ 2.5 \\
VoiceStar  & 65k  EN     & 840M  & 2.20             & 0.605             & 3.92             & & 2.14             & 0.603             & 4.25             & 60.0 $\pm$ 2.6             \\
MaskGCT    & 100k Multi. & 1048M & 2.03             & \textbf{0.712}    & 3.56             & & 2.58             & \textbf{0.691}    & 3.92             & 57.6 $\pm$ 2.6             \\
Spark-TTS  & 102k Multi. & 507M  & 1.85             & 0.572             & 3.94             & & 2.11             & 0.578             & \underline{4.35} & 63.2 $\pm$ 2.5             \\
Kyutai-TTS & 88k EN      & 750M  & 1.34             & \underline{0.689} & 3.63             & & \textbf{1.63}    & \underline{0.660} & 4.04             & 60.5 $\pm$ 2.4             \\
F5-TTS     & 100k Multi. & 336M  & \textbf{1.26}    & 0.670             & 3.69             & & 2.05             & 0.653             & 3.88             & 65.6 $\pm$ 2.4             \\
\midrule
VoXtream2-B & 40k   EN & 462M & 1.45            & 0.679             & 3.56 & & 2.14             & 0.658             & 4.02             & 60.2 $\pm$ 2.5    \\
VoXtream2  & 40k   EN    & 462M  & \underline{1.32} & 0.656             & \underline{4.05} & & 2.12             & 0.638             & 4.24             & \textbf{68.8} $\pm$ 2.3    \\
\bottomrule
\end{tabular}%
}
\vspace{-2ex}
\end{table*}

\subsection{Speaking Rate Control}

\begin{figure}[t]
  \centering
  \includegraphics[width=1\linewidth]{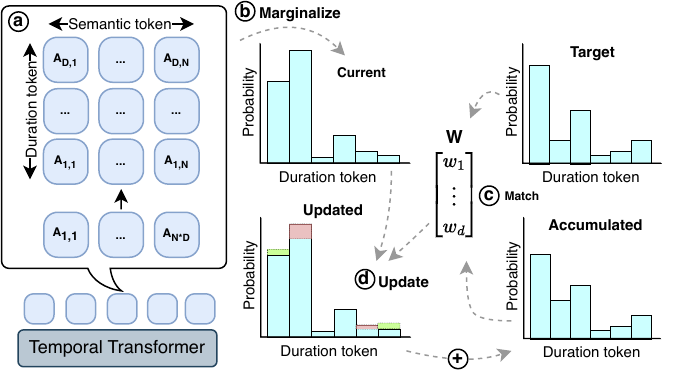}
  \caption{Speaking rate control mechanism.}
  \label{fig:spk_rate}
\vspace{-1ex}
\end{figure}

We control the speaking rate during generation using a distribution matching strategy. Phoneme-to-audio alignment provides duration tokens used for monotonic alignment in our model. For each utterance, the distribution of duration tokens forms a \textit{duration state}, represented as a 6-bin histogram where each bin corresponds to the probability of a specific duration token. Duration states associated with a given syllables-per-second (SPS) value are selected and used as an additional conditioning input during sampling of the next token ($P_{\text{target}}$).

The control mechanism is illustrated in Figure \ref{fig:spk_rate}. To obtain the current duration distribution $P_{\text{current}}$ from the TT output, we reshape the joint output vector of size $N * D$ into a $D \times N$ matrix (a) and compute the marginal distribution over $N$ (b). To enforce the target speaking rate, we compare the target duration distribution $P_{\text{target}}$ with the accumulated distribution $P_{\text{acc}}$, computed over a sliding window of previously generated segments (c),  The resulting weights are used to update the predicted duration distribution at the current step (d):

\vspace{-1ex}
\begin{gather}
P_{\text{current}} =
\mathrm{Softmax}\left(
\frac{1}{T}\log\sum_{n=1}^{N}\exp(A_{dn})
\right)
\\[6pt]
W =
\exp\Big(
\beta\big(
\log_{10}(P_{\text{target}})
-
\log_{10}(P_{\text{acc}})
\big)
\Big)
\\[6pt]
P_{\text{updated}} =
\frac{
P_{\text{current}} \odot W
}{
\sum_{i=1}^{D} P_{\text{current},i} W_i
}
\end{gather}

Here, $T=0.9$ is the sampling temperature, and $A_{dn}$ denotes the TT output for duration $d$ and semantic token $n$. $P_{\text{acc}}$ is initialized as a uniform distribution and estimated from a duration counter over the past 3 seconds of generated speech, which is long enough for a robust estimate and short enough to react to changes. The updated distribution $P_{\text{updated}}$ is used to sample the next duration token via nucleus sampling (top-p = 0.9). The parameter $\beta$ controls the strength of SRC: with $\beta=1$, the effect on intelligibility is minimal, but control is weak. With $\beta=10$, the speaking rate follows the target more closely, although WER increases due to hallucinations. We therefore set $\beta=5$ as a trade-off between quality and intelligibility. All bins in $P_{\text{target}}$ and $P_{\text{acc}}$ are strictly positive after smoothing.

The selected duration token determines the corresponding row in matrix (a), after which we apply top-k sampling over semantic tokens with $k=5$. Acoustic tokens from DT are generated using greedy sampling for consistency. Distribution matching is optional and applied only when SRC is enabled. Otherwise, duration tokens are sampled directly from $P_{\text{current}}$. To avoid a speaking-rate increase observed in Parakeet \cite{darefsky2024parakeet} and SSR-Speech \cite{wang2025ssr}, we do not apply CFG to the duration state.

To improve naturalness at slow speaking rates, we encourage implicit filler-word generation during training. Instead of providing explicit cues, the model must infer filler placement from prosodic patterns, primarily elongated phoneme durations. Consequently, it learns to insert fillers automatically according to speaking rate and sentence context.

\section{Experiment Setup}

\subsection{Datasets}
\label{sec:data}
We used the Emilia \cite{he2024emilia} spontaneous speech dataset and HiFiTTS-2 \cite{langman25_interspeech}, derived from LibriVox \cite{kearns2014librivox} audiobooks, as the basis of our training corpus. From Emilia, we selected the 47k-hour English subset. From HiFiTTS-2, we used a 15k-hour 22kHz training subset, removing utterances shorter than 5 seconds or with a word error rate (WER) above 10\%. IPA phonemes were extracted from transcripts using espeak-ng phonemizer\footnote{\url{https://github.com/espeak-ng/espeak-ng}}. Phoneme-to-audio alignment was performed with the Clap-IPA forced aligner \cite{zhu2024taste}. After filtering failed alignments and invalid transcripts, the final corpus comprised 30k hours from Emilia and 10k hours from HiFiTTS-2, totaling 40k hours. For speech tokenization, we used the Mimi \cite{defossez2024moshi} audio codec at 24\,kHz. Filler words (e.g., ``uh'', ``uhm'', ``yeah'') were removed from transcripts, and their phoneme timestamps were merged with the preceding phoneme in the alignment (or the following phoneme if the sentence began with filler). To reach a target duration of 55 seconds per sample, we concatenated utterances at speaker level and padded shorter clips with silence.

\subsection{Model}

Following VoXtream \cite{torgashov2026voxtream}, we adopt an open-source implementation of the Llama-3.2 \cite{dubey2024llama} transformer as the backbone of our model. The TT module has 12 layers, 16 attention heads, an embedding size of 1024, and a feed-forward dimension of 4096. The PT consists of 6 layers with 8 attention heads. The DT contains 4 layers, 8 attention heads, and a feed-forward dimension of 8192, following the Sesame-CSM model \cite{sesame2025uncannyvoice}. We keep the DT weights frozen, as it was pretrained on a large-scale conversational speech dataset\footnote{\url{https://github.com/SesameAILabs/csm}}. Speaker embeddings are extracted using a ReDimNet \cite{yakovlev24_interspeech} trained on over 100k identities.

Training was performed for 28 hours on 2$\times$NVIDIA H200 GPUs with a batch size of 64 per GPU for 10 epochs. During training, we randomly crop fixed 50\,s audio segments. We use the AdamW optimizer \cite{loshchilov2017decoupled}, linearly warming up the learning rate during the first epoch from $1 \times 10^{-5}$ to $2 \times 10^{-4}$. The training graph is compiled with \texttt{torch.compile} to maximize GPU utilization. During generation, TT and DT are wrapped with CUDA Graphs following Moshi \cite{defossez2024moshi}, and the streaming state of the Mimi codec is cached, significantly improving performance over the original VoXtream. CFG is implemented via batching with negligible overhead. All random seeds are fixed during training and generation to ensure reproducibility.

\subsection{Baseline models}

We selected publicly available zero-shot TTS models with full-stream or duration-control capabilities as baselines. MaskGCT \cite{wang2024maskgct} and F5-TTS \cite{chen2025f5} are non-autoregressive (NAR) models, while VoiceStar \cite{peng2025voicestar} and Spark-TTS \cite{wang2025spark} are autoregressive (AR) models with explicit duration control. CosyVoice2 \cite{du2024cosyvoice2}, Kyutai-TTS \cite{kyutai2025streaming}, and VoXtream \cite{torgashov2026voxtream} are AR full-stream models. CosyVoice2 additionally supports SRC via instructed generation. For a fair comparison, we restricted models to those with up to 1B parameters and training data not exceeding 200k hours, prioritizing systems trained on public datasets such as Emilia and LibriVox. All results were reproduced using official implementations and publicly available checkpoints.

\subsection{Evaluation}

\label{sec:eval}

We evaluate VoXtream2 on three test sets. The first is the LibriSpeech-PC \cite{meister2023librispeech} \textit{test-clean} subset, where we follow the cross-sentence protocol introduced in F5-TTS \cite{chen2025f5}. The second is SEED-TTS \cite{anastassiou2024seed} \textit{test-en}, evaluated using the original protocol.

The third set targets SRC evaluation and is derived from the Emilia-EN \cite{he2024emilia} dataset, which we denote as Emilia \textit{speaking-rate}. We selected 62 speakers (balanced by gender), half of whom include filler words in the prompt (uniformly distributed across genders). Each speaker provides three audio prompts with slow ($2.0\pm0.5$ SPS), normal ($4.0\pm0.6$ SPS), and fast ($5.6\pm0.5$ SPS) speaking rates. The prompt length distribution is similar to SEED \textit{test-en}, with a mean of 4.5 seconds. Texts were generated using ChatGPT \cite{openai_gpt5_2025}, prompted with conversational-style sentences from the selected speakers and instructed to produce stylistically similar sentences containing 30-40 syllables (approximately 9-10 seconds of speech at a moderate speaking rate). For dynamic SRC evaluation, we used the same procedure but requested longer sentences of 50--80 syllables to capture multiple speaking-rate transitions within a single utterance. All speakers were unseen during the training of VoXtream2. The dataset is available at \href{https://huggingface.co/datasets/herimor/voxtream2-test/}{hf.co/datasets/herimor/voxtream2-test/}.

We report three reproducible objective metrics. Intelligibility is measured by WER between the transcription of synthesized speech and the input text. We use Whisper-large-v3 ASR \cite{radford2023robust} for SEED-TTS \textit{test-en} and Emilia \textit{speaking-rate}, and a HuBERT-based ASR \cite{hsu2021hubert} for LibriSpeech-PC \textit{test-clean}. Following \cite{kyutai2025streaming}, we apply the Whisper text normalizer before computing WER. Speaker similarity (SPK-SIM) is computed as the cosine similarity between embeddings extracted from the acoustic prompt and synthesized speech using a WavLM-based \cite{chen2022wavlm} ECAPA-TDNN \cite{desplanques20_interspeech} model. Signal quality is monitored with the UTMOS predictor \cite{saeki22c_interspeech}.

For naturalness evaluation, we conducted two user studies on the Prolific platform: one for short-form TTS and one for controllable speaking rate. We sampled 100 unique sentences with WER below 10\% to exclude hallucinations. In each study, 40 native listeners rated the naturalness of each model on a 0--100 scale using a MUSHRA-like protocol. Multiple attention checks were included, and participants were compensated at an average rate of 9 GBP per hour.

To evaluate static SRC, we visualized the correlation between target and generated speaking rates. We also analyzed filler insertion frequency as a function of speaking rate and prompt context, and examined the influence of prompt speaking rate on the output. For dynamic SRC, we report the Pearson correlation (Corr.) to quantify how well the system follows rate-control instructions over time. In the full-stream setting, input text is provided word by word. We measure the real-time factor (RTF), defined as the ratio of generated speech duration to wall-clock time, and first-packet latency (FPL), the time to the first speech frame. Both metrics are measured from text input to waveform output. We use the first 10 rows of the SEED \textit{test-en} protocol as a benchmark, with one additional utterance for warmup. To simulate LLM-style streaming, we vary the input text rate to 10, 20, and 40 tokens per second and evaluate robustness under these conditions. For simplicity, we consider a single word as a token in that evaluation.

\section{Results}

\begin{table}[t]
  \caption{Evaluation of prompt text masking on different prompt speaking rates. Mean $\pm$ STD is reported.}
  \label{tab:prompt_spk_rate}
  \centering
  \resizebox{0.95\linewidth}{!}{
  \begin{tabular}{lccc}
    \toprule
    \textbf{Model} & \textbf{Slow} & \textbf{Normal} & \textbf{Fast} \\
    \midrule
    \multicolumn{4}{c}{\textbf{WER (\%) $\downarrow$}} \\
    \midrule
    Baseline
      & $2.80 \pm 5.64$
      & $2.10 \pm 4.32$
      & $4.21 \pm 8.10$ \\
    Prompt mask
      & $2.14 \pm 3.27$
      & $2.20 \pm 3.54$
      & $2.13 \pm 3.73$ \\
    \midrule
    \multicolumn{4}{c}{\textbf{Speaking rate}} \\
    \midrule
    Baseline
      & $3.15 \pm 0.82$
      & $3.48 \pm 0.80$
      & $3.58 \pm 0.84$ \\
    Prompt mask
      & $2.76 \pm 0.72$
      & $3.09 \pm 0.74$
      & $3.37 \pm 0.84$ \\
    \bottomrule
  \end{tabular}
}
\vspace{-2ex}
\end{table}

In Table~\ref{tab:eval}, we compare VoXtream2 with prior zero-shot TTS systems that support streaming, duration control, or both. Despite its smaller model size and substantially less training data, VoXtream2 achieves performance comparable to state-of-the-art systems. In subjective evaluation, VoXtream2-B is on par with Kyutai-TTS and VoiceStar, and becomes the most preferred system with prompt enhancement. Listening to the samples suggests that most systems exhibit strong prosody and pronunciation, with perceived differences mainly driven by audio quality, explaining both the preference gain from prompt enhancement and the lower score of the human recordings, which contain real background noise and artifacts. This observation aligns with prior findings that MOS ratings are sensitive to evaluation design and may reflect overall quality rather than strictly naturalness~\cite{kirkland2023stuck}. The original VoXtream is excluded from this comparison due to its much smaller training corpus.

\begin{table}[t]
\caption{Performance in mixed precision on an RTX3090 GPU in full-stream. \textbf{TC} denotes Torch Compile, \textbf{CG} denotes CUDA graphs, and \textbf{TRT} corresponds to TensorRT.}
\label{tab:rtf}
\centering

\resizebox{0.7\linewidth}{!}{
\begin{tabular}{lcc}
\toprule
  \textbf{Model} &
  \textbf{FPL(ms) $\downarrow$} &
  \textbf{RTF} $\downarrow$ \\
\midrule
CosyVoice2:\textbf{TRT} & 837         & 0.546          \\
Kyutai-TTS:\textbf{CG}  & 277         & 0.272          \\
VoXtream:\textbf{TC}    & 78          & 0.223          \\
\midrule
VoXtream2:\textbf{CG}  & 74          & 0.256          \\
VoXtream2:\textbf{TC}   & \textbf{63} & \textbf{0.173} \\
\bottomrule
\end{tabular}
}
\end{table}

\begin{figure}[t]
  \centering
  \hspace{-2ex}
  \includegraphics[width=\linewidth]{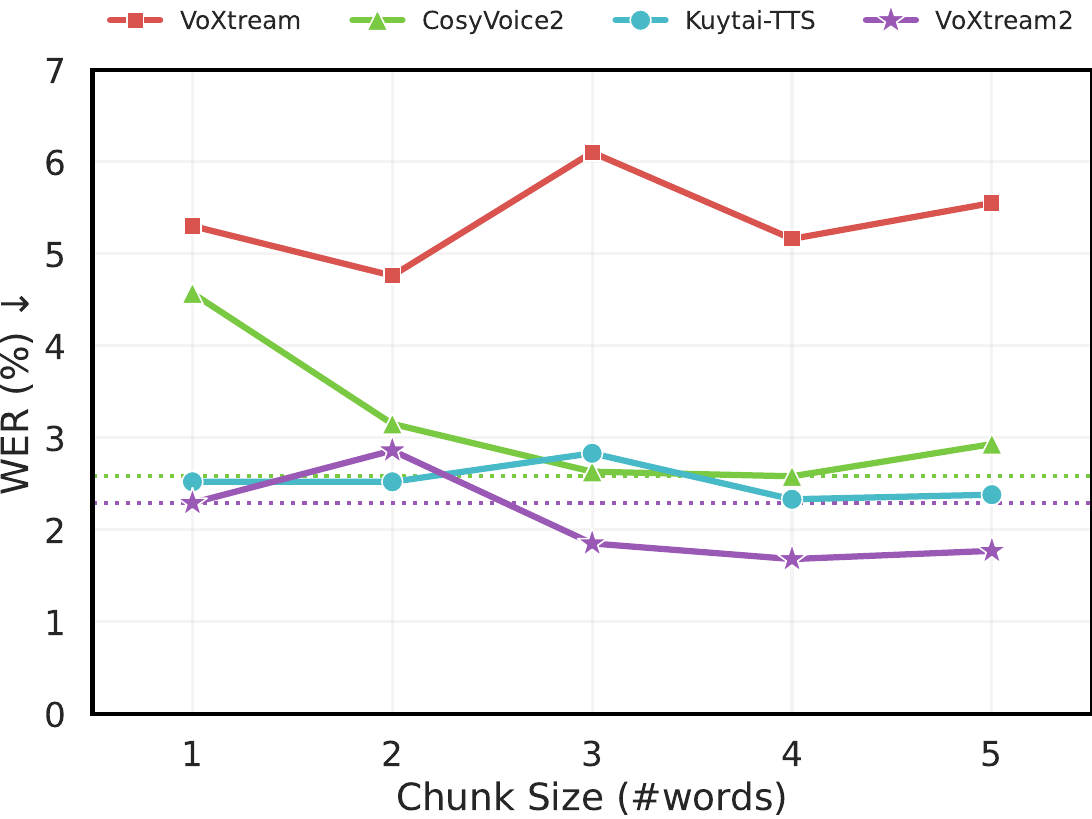}
  \caption{Evaluation of text chunk size in a full-stream.}
  \label{fig:full_stream_wer}
  \vspace{-2ex}
\end{figure}

Table \ref{tab:prompt_spk_rate} presents a comparison between models trained with and without prompt text masking on the Emilia \textit{speaking-rate} dataset. The proposed approach removes the need for an external phoneme aligner while maintaining stable WER across different prompt speaking rates relative to the baseline. However, the generated speaking rate remains influenced by the prompt rate, even with prompt text masking. For fairness, neither CFG nor SRC was applied in this experiment.

\subsection{Full-stream}

We benchmark the performance of our model against prior work in Table \ref{tab:rtf}. VoXtream2 achieves the lowest FPL and RTF among publicly available full-stream models. We also evaluate different GPU inference optimizations. While \texttt{torch.compile} provides the best runtime performance, it requires longer graph compilation time compared to CUDA Graphs.

We further analyze how input text chunk size affects intelligibility in the full-stream setting on the Emilia \textit{speaking-rate} dataset (Figure \ref{fig:full_stream_wer}). VoXtream shows only minor improvement at chunk size 2 due to its limited look-ahead. Its overall WER is higher than that of other models, partly because of its reliance on an external phoneme aligner, which struggles with fast-speaking prompts. CosyVoice2 exhibits substantial WER improvement as chunk size increases. However, even at chunk size 4, it remains less accurate than VoXtream2. KyutaiTTS achieves WER comparable to VoXtream2 under word-level streaming (chunk size 1), but benefits only marginally from additional context. In contrast, VoXtream2 attains the lowest WER for word-level streaming and shows consistent gains as chunk size increases. We do not report SPK-SIM or UTMOS for the full-stream setting, as these metrics show no significant differences compared to non-streaming evaluation.

\begin{figure}[!t]
  \centering
  \includegraphics[width=1.05\linewidth]{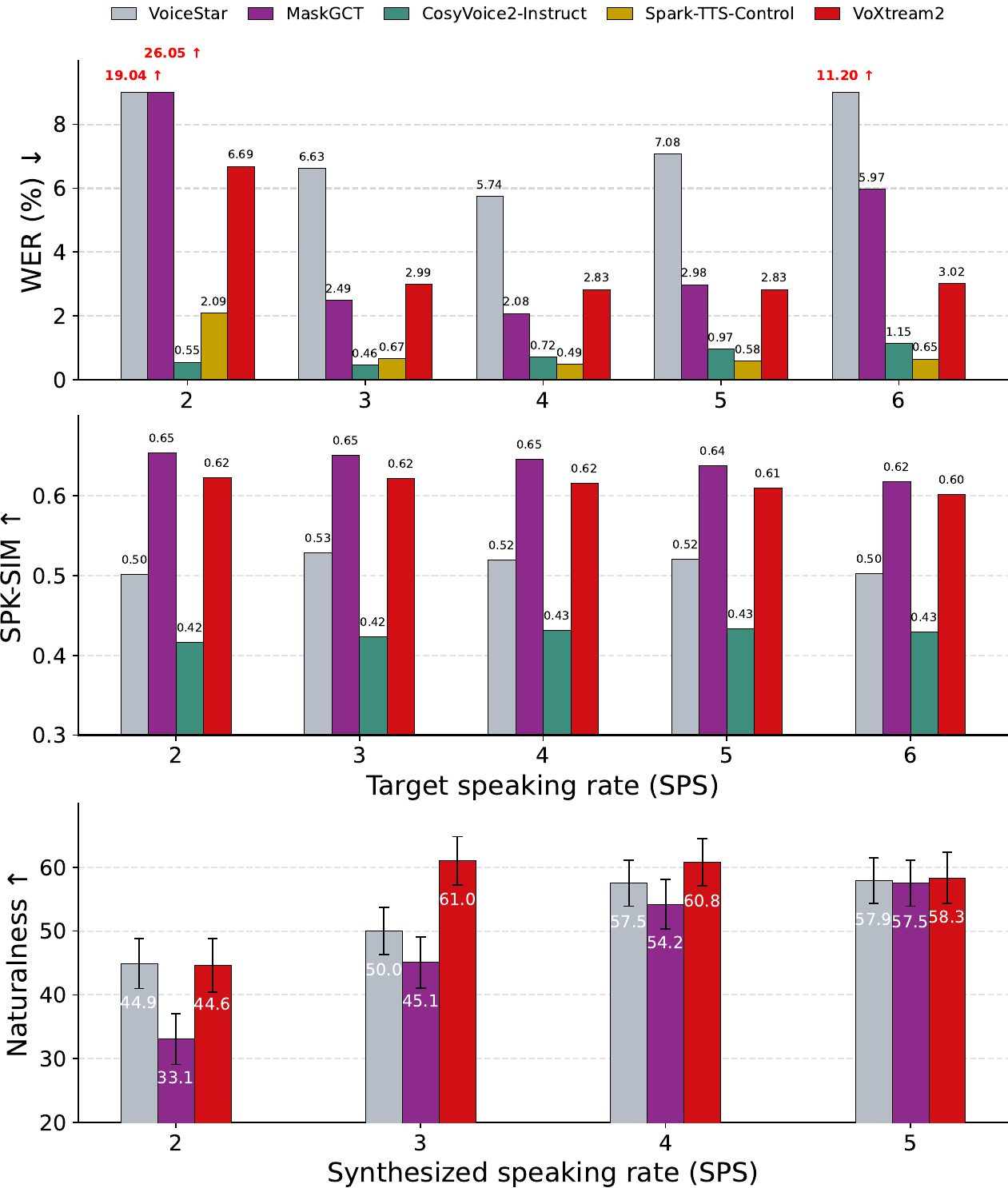}
  \caption{Comparison of different TTS models across various speaking rates for utterance-level control.}
  \label{fig:wer_and_spk_sim_by_model}

  \vspace{0.6em}

  \includegraphics[width=0.73\linewidth]{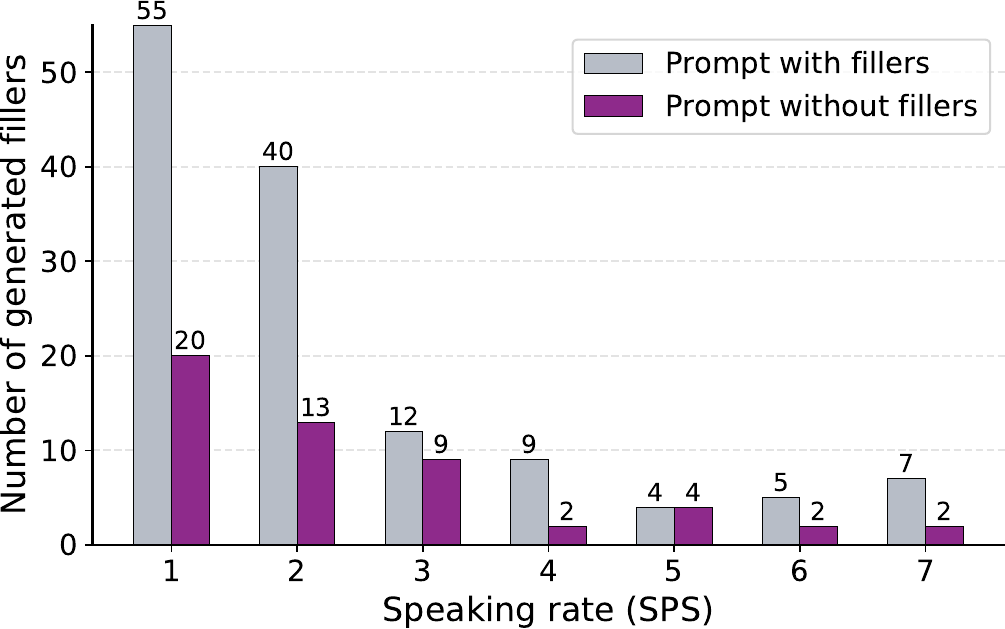}
  \caption{Dependency between the number of generated filler words and speaking rate.}
  \label{fig:fillers_cnt_group}
\vspace{-3ex}
\end{figure}

\begin{figure}[!t]
  \centering
  \includegraphics[width=\linewidth]{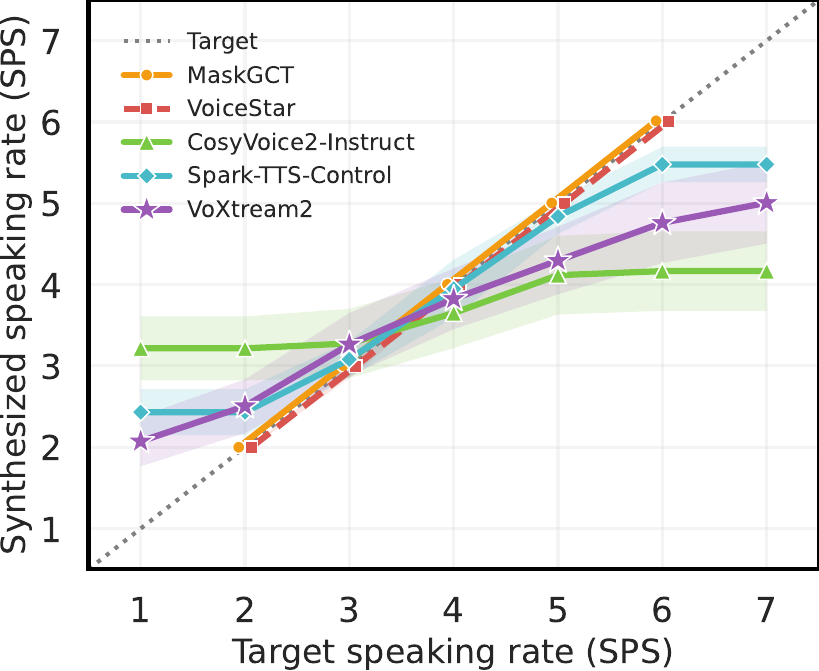}
  \caption{Correlation between target and synthesized speaking rates across different TTS models.}
  \label{fig:spk_rate_by_model}

  \vspace{0.6em}

  \includegraphics[width=\linewidth]{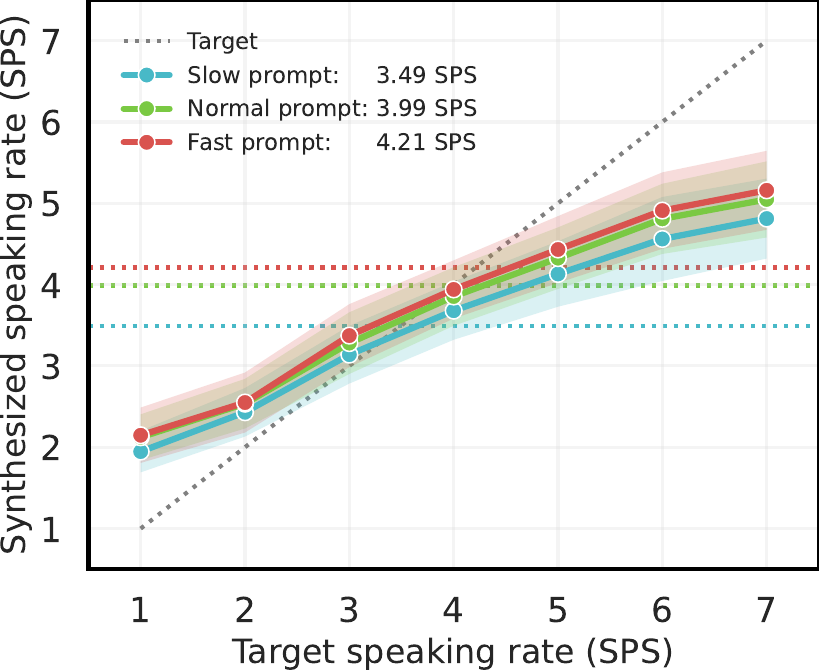}
  \caption{Influence of the prompt speaking rate on the predicted speaking rate.}
  \label{fig:prompt_spk_rate_vs_pred_spk_rate}
  \vspace{-2ex}
\end{figure}

\subsection{Static speaking rate control}

We use the Emilia \textit{speaking-rate} dataset for all SRC evaluations. Figure \ref{fig:spk_rate_by_model} presents the correlation between generated and target speaking rates. CosyVoice2 with instructed generation adjusts the speaking rate only within a limited range. In contrast, VoXtream2 and Spark-TTS demonstrate substantially stronger controllability. Spark-TTS provides more accurate control for fast speech, but supports only five discrete speaking-rate states, whereas VoXtream2 allows continuous control within the 2--5 SPS range. VoiceStar and MaskGCT enable precise duration control and closely match the target speaking rate, though at the cost of degraded speech quality. We exclude F5-TTS from this comparison, as we were unable to generate intelligible speech with strict duration control. Duration-control models are evaluated only within the 2--6 SPS range, consistent with the operating range of other systems.

Objective results for static SRC on Emilia \textit{speaking-rate} are shown in the first two plots of Figure \ref{fig:wer_and_spk_sim_by_model}. VoXtream2 yields higher WER than CosyVoice2 and Spark-TTS, but provides significantly stronger voice-cloning performance. MaskGCT achieves comparable WER at moderate rates (3 and 5 SPS) and performs notably better at 4 SPS. However, its intelligibility drops sharply at extreme rates. In terms of speaker similarity, MaskGCT slightly outperforms VoXtream2 across evaluated rates. VoiceStar exhibits the highest WER and similar degradation at extreme speaking rates. Although its SPK-SIM remains stable, it is lower than that of VoXtream2. Spark-TTS is excluded from SPK-SIM evaluation because its controllable version does not support voice cloning (average SPK-SIM=0.05).

Subjective results are shown in the third plot of Figure \ref{fig:wer_and_spk_sim_by_model}. For fairness, we exclude Spark-TTS (no voice cloning) and CosyVoice2 (limited to 3--4 SPS). For VoXtream2, samples were selected according to the achieved speaking rate (e.g., target rate 1 for 2 SPS and 7 for 5 SPS). VoXtream2 significantly outperforms other systems at 3 SPS and is comparable at 5 SPS. At 4 SPS, it achieves slightly higher preference scores, and at 2 SPS, it performs on par with VoiceStar. For slow speech (2--3 SPS), VoXtream2 clearly outperforms MaskGCT.

To analyze the influence of prompt speaking rate, we compute the correlation between generated and target rates for prompts with different speaking speeds. Figure \ref{fig:prompt_spk_rate_vs_pred_spk_rate} shows that SRC substantially reduces dependence on the prompt rate. The dotted curves indicate generation without SRC. With SRC enabled, the achievable speaking-rate range is significantly broader than the variation induced by ``slow,'' ``normal,'' and ``fast'' prompts, and differences between curves are much smaller than those observed without SRC.

We further analyze filler-word generation (Figure \ref{fig:fillers_cnt_group}). A clear relationship is observed between filler frequency and speaking rate: fillers are more frequent at slow rates and nearly disappear at fast rates, consistent with human speech. Prompts containing fillers further increase their occurrence. However, even without such cues, VoXtream2 inserts fillers automatically at slow rates. Among the evaluated systems, only VoXtream2 exhibits this adaptive filler insertion behavior.


\subsection{Dynamic speaking rate control}

\begin{table}[t]
  \caption{Evaluation of dynamic speaking rate control. S, N, and F denote Slow, Normal, and Fast target speaking rates.}
  \label{tab:dynamic_spk_rate}
  \centering
  \resizebox{\linewidth}{!}{
  \begin{tabular}{lcccc}
    \toprule
    \textbf{Model} &
    \textbf{Corr.} &
    \textbf{WER (\%) $\downarrow$} &
    \textbf{SPK-SIM $\uparrow$} &
    \textbf{UTMOS $\uparrow$} \\
    \midrule
    No SRC                            & -     & 3.26  & 0.624 & 3.86 \\
    S (1 SPS)                         & -     & 16.51 & 0.620 & 3.53 \\
    N (4 SPS)                         & -     & 3.20  & 0.626 & 3.83 \\
    F (7 SPS)                         & -     & 3.52  & 0.609 & 3.81 \\
    \midrule
    S $\rightarrow$ N $\rightarrow$ F & 0.701 & 4.73  & 0.628 & 3.75 \\
    F $\rightarrow$ N $\rightarrow$ S & 0.826 & 11.25 & 0.626 & 3.71 \\
    S $\leftrightarrow$ F             & 0.619 & 9.55  & 0.629 & 3.66 \\
    F $\leftrightarrow$ S             & 0.660 & 11.20 & 0.627 & 3.64 \\
    \bottomrule
  \end{tabular}
}
\end{table}


\begin{figure}[t]
  \centering
  \hspace*{-0.35cm}
  \includegraphics[width=1\linewidth]{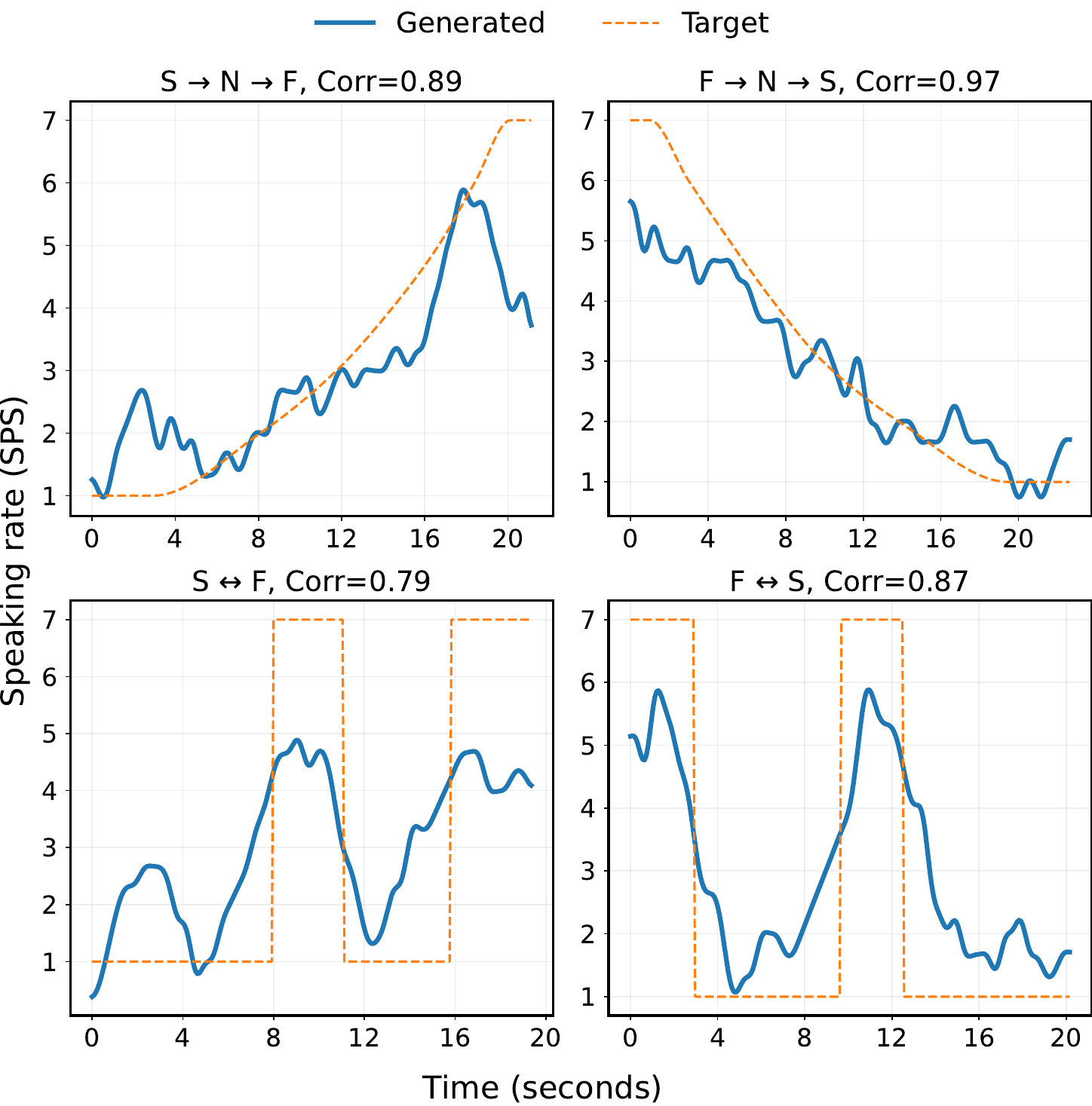}
  \caption{Examples of dynamic speaking rate control. S, N, and F denote Slow, Normal, and Fast target speaking rates.}
  \label{fig:dynamic_spk_rate}
  \vspace{-2ex}
\end{figure}

To evaluate dynamic speaking-rate control, we conduct experiments on the Emilia \textit{speaking-rate} test set and compute objective metrics. We use the same acoustic prompts as in the static evaluation, but generate longer sentences (50--80 syllables). Four scenarios are considered. The first two assess gradual rate changes by linearly increasing or decreasing the SRC control signal within the 1--7 SPS range. The other two are more challenging: we alternate between slow (1 SPS) and fast (7 SPS) target rates every 40 phonemes. To obtain the speaking-rate curve for each utterance, we extract phoneme alignments from our model, convert phonemes to syllables, and compute SPS over a 3-second sliding window with 25\% overlap. The curve is then interpolated to the time axis. For reference, we also report static SRC results for the same SPS range.

In the reference static SRC setting (Table \ref{tab:dynamic_spk_rate}), the model behaves consistently at a normal speaking rate. At a fast rate, SPK-SIM decreases slightly while WER remains nearly unchanged. At a slow rate, however, WER increases substantially. Manual inspection reveals two main causes: hallucinations due to the underrepresentation of very slow duration states in the training data, and increased word repetition or filler insertion. Repetition and filler insertion can make the speech sound more human-like, but they additionally inflate the WER.

Results in Table \ref{tab:dynamic_spk_rate} show that the generated speaking rate correlates well with the control signal. Compared to the baseline, WER increases at extreme control values as hallucinations become more frequent, whereas SPK-SIM remains stable and UTMOS degrades only slightly. The Corr.\ metric indicates that starting with a fast speaking rate yields stronger controllability than starting slowly, although it negatively affects intelligibility.


Figure \ref{fig:dynamic_spk_rate} illustrates examples of dynamic SRC. The model follows gradual changes and adapts rapidly to abrupt transitions in the control signal. Short-term fluctuations around the target rate resemble natural speaking-rate variation in human speech and help maintain naturalness rather than indicating control failure. The system reliably generates speech at 1 SPS, while the maximum reliable speaking rate is around 5--6 SPS.

\section{Ablations}

Table \ref{tab:arch_ablation} presents a modular ablation analysis of VoXtream2 on the SEED \textit{test-en} dataset. The first row shows improvements across all objective metrics compared to VoXtream, achieved through dataset scaling and architectural changes. The second row demonstrates a significant reduction in WER after applying CFG only to the text conditioning. The following row shows a notable improvement in SPK-SIM after introducing audio CFG, although this comes at the cost of higher WER and lower UTMOS. Adding speaker-embedding CFG further improves SPK-SIM and slightly reduces WER, but again results in a small drop in UTMOS. Increasing the weight of speaker-embedding conditioning in the penultimate row has a similar effect and yields the best SPK-SIM score. The final row incorporates a speech enhancement module for the acoustic prompt, achieving the best UTMOS and WER, with a slight reduction in SPK-SIM.

\begin{table}[t]
  \caption{Modular analysis of VoXtream2 modifications.}
  \label{tab:arch_ablation}
  \centering
  \hspace*{-0.2cm}%
  \resizebox{1.05\linewidth}{!}{
  \begin{tabular}{lccc}
    \toprule
    \textbf{Model} &
    \textbf{WER (\%) $\downarrow$} &
    \textbf{SPK-SIM $\uparrow$} &
    \textbf{UTMOS $\uparrow$} \\
    \midrule
    VoXtream \cite{torgashov2026voxtream} & 3.16 & 0.529 & 3.88 \\
    \midrule
     \hspace{0.1cm}+ Data \& updates      & 2.29             & 0.578             & \underline{3.94} \\
     \hspace{0.2cm}+ CFG text             & 1.37             & 0.578             & 3.91             \\
     \hspace{0.3cm}+ CFG audio            & 1.62             & 0.661             & 3.65             \\
     \hspace{0.4cm}+ CFG speaker          & 1.55             & \underline{0.674} & 3.60             \\
     \hspace{0.5cm}+ Speaker weight       & \underline{1.45} & \textbf{0.679}    & 3.56             \\
     \hspace{0.6cm}+ Prompt enhance. & \multirow{2}{*}{\textbf{1.32}} & \multirow{2}{*}{0.656} & \multirow{2}{*}{\textbf{4.05}} \\
     \hspace{0.9cm} (VoXtream2) & & & \\
    \bottomrule
  \end{tabular}
}
\end{table}

\begin{table}[t]
\caption{Effect of input text streaming rate (TPS) and phoneme look-ahead (LA) on full-stream TTS performance.}
\label{tab:delayed_text}
\centering

\resizebox{0.9\linewidth}{!}{
\begin{tabular}{ccccc}
\toprule
  \textbf{TPS} &
  \textbf{LA} &
  \textbf{WER (\%) $\downarrow$} &
  \textbf{FPL(ms) $\downarrow$} &
  \textbf{RTF} $\downarrow$ \\
\midrule
-  & $\infty$ & 1.44 & 44  & 0.229 \\
\midrule
$\infty$ & 3        & 2.29 & 75  & 0.256 \\
40       & 3        & 2.29 & 75  & 0.254 \\
20       & 3        & 2.98 & 67  & 0.259 \\
\midrule
10      & 1       & 56.13 & 56  & 0.262  \\
10      & 2       & 22.89 & 56  & 0.274  \\
10      & 3       & 5.11  & 81  & 0.274  \\
10      & 4       & 3.46  & 146 & 0.271  \\
10      & 5       & 2.66  & 176 & 0.280  \\
\bottomrule
\end{tabular}
}
\vspace{-2ex}
\end{table}

\begin{figure}[t]
  \centering
  \includegraphics[width=\linewidth]{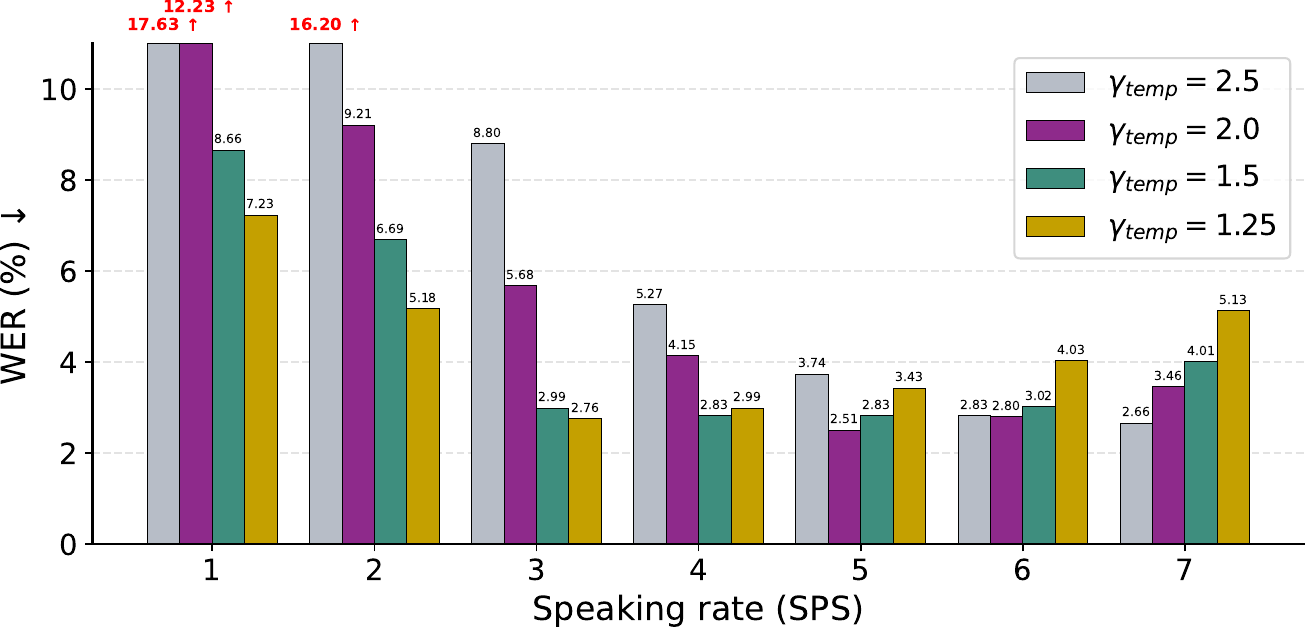}
  \caption{Influence of the CFG $\gamma$ on the speaking rate control.}
  \label{fig:spk_rate_cfg_wer}
  \vspace{-1ex}
\end{figure}

Motivated by the speaking-rate acceleration effect of CFG observed in \cite{darefsky2024parakeet,wang2025ssr}, we investigated the influence of the $\gamma_{temp}$ parameter on VoXtream2 under SRC. Since CFG is not applied during duration-token sampling, we did not observe a significant change in speaking rate. However, we found a clear relationship between $\gamma_{temp}$ and WER in SRC (Figure \ref{fig:spk_rate_cfg_wer}). Lower $\gamma_{temp}$ values yield the best WER for slow speech but the worst for fast speech, while larger values produce the lowest WER for fast speech but degrade intelligibility at slow rates. These results indicate that CFG still indirectly influences speaking-rate behavior even without being applied to duration states. Under SRC, this effect mainly appears as changes in intelligibility rather than speed. This suggests that WER for SRC could be further improved by tuning $\gamma_{temp}$ for the target speaking rate.

To evaluate robustness when receiving text from an upstream LLM, we varied the tokens-per-second (TPS) rate of the input text stream and measured its impact on performance. The look-ahead (LA) column indicates how many future phonemes are required to generate the current audio frame. Infinite TPS corresponds to immediate text availability with no delay. Results are summarized in Table \ref{tab:delayed_text}. The first row reports metrics for output streaming to quantify the increase in FPL and WER in the full-stream setting. The next rows show that TPS=40 (25\,ms per token, similar to \cite{sheng2025syncspeech}) does not affect the model, as speech generation remains slower than text arrival, effectively providing ample look-ahead. At TPS=20, WER increases because the TTS model and text generator operate at similar speeds, reducing effective look-ahead. In the final rows, TPS is further reduced so text arrives slower than speech generation. We then analyze the effect of LA on WER and generation speed. With LA below 3, FPL is minimal but the output becomes unintelligible. LA=3 provides the best trade-off, maintaining moderate WER even at TPS=10 while keeping FPL low. Increasing LA to 5 phonemes significantly reduces WER but more than doubles FPL, illustrating the latency–quality trade-off.

\section{Limitations}

VoXtream2 is not fully disentangled from the speaking rate of the acoustic prompt. As shown in Table \ref{tab:prompt_spk_rate}, the generated speaking rate is still influenced by the prompt rate. Although the proposed SRC method (Figure \ref{fig:prompt_spk_rate_vs_pred_spk_rate}) partially mitigates this effect, we observe an increase in WER for certain combinations of prompt and target speaking rates. For instance, prompts with a slow speaking rate tend to increase WER when generating fast speech, while fast prompts degrade WER for slow targets. This behavior arises because the acoustic prompt is encoded using the same model that generates speech tokens, which is trained under a next-token prediction objective and therefore tends to continue the sequence in a similar speaking style. Future work could explore using a dedicated prompt encoder or alternative training objectives to better disentangle prompt and target speaking rates.

Another limitation of VoXtream2 is the complexity of the data preprocessing pipeline. Although the external phoneme aligner is not required during generation, it remains a crucial component of training data preparation. As shown in Section \ref{sec:data}, approximately 35\% of the data had to be discarded due to invalid alignments. In future work, we plan to explore alternative alignment methods to reduce this dependency.

\section{Conclusion}

VoXtream2 shows that ultra-low-latency full-stream zero-shot TTS can be paired with practical, fine-grained control. The proposed dynamic speaking rate control can be updated mid-utterance, enabling smooth transitions between slow, normal, and fast speech while maintaining stable intelligibility and voice cloning in the operating range. Prompt text masking enables textless prompts, removing a major usability barrier and reducing sensitivity to prompt content and speaking rate. Classifier-free guidance applied across conditioning signals improves intelligibility and speaker similarity in zero-shot TTS and provides an effective lever to trade off similarity, signal quality, and controllability. Moreover, while CFG is not applied to duration sampling, varying $\gamma_{temp}$ still affects SRC outcomes by shifting intelligibility across target rates (favoring slow speech at lower $\gamma_{temp}$ and fast speech at higher $\gamma_{temp}$). Finally, we find that VoXtream2 remains robust under realistic LLM-driven text streams: at moderate token rates, the effective look-ahead is sufficient to preserve quality, whereas overly constrained look-ahead yields the expected latency--quality trade-off. Overall, evaluations indicate that VoXtream2 is competitive with publicly available zero-shot baselines and significantly faster than real-time on user-grade hardware, making it suitable for real-time conversational applications.

\ifcameraready
\section{Acknowledgements}
\label{sec:ackn}

This work was partially supported by the Wallenberg AI, Autonomous Systems and Software Program (WASP) funded by the Knut and Alice Wallenberg Foundation (KAW) and by the Industrial Strategic Technology Development Program (grant no.\ 20023495) funded by MOTIE, South Korea. Computations were enabled by the Berzelius resource provided by KAW at the National Supercomputer Centre.
\else
\fi

\section{GenAI Usage Disclosure}
ChatGPT \cite{openai_gpt5_2025} was used to generate sentences for the Emilia \textit{speaking-rate} test, coding the web interface for user studies and polishing of the manuscript. All AI-generated code and text were thoroughly verified by the authors.

\bibliographystyle{IEEEtran}
\bibliography{main}

\end{document}